\documentclass{natureprintstyle}

\spacing{1}
\spacing{2}
\spacing{1}

\usepackage{amsmath,amssymb}
\usepackage[applemac]{inputenc}
\usepackage{graphicx}
\usepackage{color}
\usepackage{bm}
\usepackage{longtable}
\usepackage{amssymb}
\usepackage{rotating}
\usepackage{hyperref}

\usepackage{hyperref}

\let\citep\cite
\let\citet\cite

\title{\sffamily Precision Cosmology from Future Lensed Gravitational Wave and Electromagnetic Signals}

\author{\sffamily Kai Liao$^{1,2}$, Xi-Long Fan$^3$, Xuheng Ding$^{1,4,5}$, Marek Biesiada$^{4,6}$, Zong-Hong Zhu$^{1,4}$}

\begin{document}
\maketitle

\begin{affiliations}
\item School of Physics and Technology, Wuhan University, Wuhan 430072, China
\item School of Science, Wuhan University of Technology, Wuhan 430070, China
\item Department of Physics and Mechanical and Electrical Engineering, Hubei University of Education, Wuhan 430205, China
\item Department of Astronomy, Beijing Normal University, Beijing 100875, China
\item Department of Physics and Astronomy, University of California, Los Angeles, CA, 90095-1547, USA
\item Department of Astrophysics and Cosmology, Institute of Physics, University of Silesia, Uniwersytecka 4, 40-007 Katowice, Poland
\end{affiliations}

\let\thefootnote\relax\footnote{
Corresponding authors: Z.-H. Zhu: zhuzh@whu.edu.cn; X.-L. Fan: fanxilong@outlook.com
}

\vspace{-3.5mm}
\begin{abstract}
\sffamily
The standard siren approach of gravitational wave cosmology appeals to the direct luminosity distance estimation
through the waveform signals from inspiralling double compact binaries, especially those with electromagnetic counterparts providing redshifts.
It is limited by the calibration uncertainties in strain amplitude and relies on the fine details of the waveform.
The Einstein Telescope is expected to produce $\mathbf{10^4-10^5}$ gravitational wave detections per year, $\mathbf{50-100}$ of which will be lensed.
Here we report a waveform-independent strategy to achieve precise cosmography by combining the accurately measured time delays from strongly lensed gravitational wave signals with the images and redshifts observed in the electromagnetic domain. We demonstrate that just 10 such systems can provide a Hubble constant uncertainty of $\mathbf{0.68\%}$ for a flat
Lambda Cold Dark Matter universe in the era of third generation ground-based detectors.
\end{abstract}

\sffamily

\begin{center}
{\textbf{ \Large \uppercase{Introduction}} }
\end{center}

The incoming era of precision cosmology requires not only more accurate but also independent probes of the Universe.
So far, however, all the information about the Universe was carried by electromagnetic (EM) waves.
Currently a tension exists between Planck satellite measurements of the Cosmic Microwave Background (CMB)~\cite{Planck} and its inferred Hubble constant ($H_{\mathrm{0}}$, which sets the present-day expansion
rate as well as the size, density and age of our Universe) and direct measurements of $H_{\mathrm{0}}$ based on the cosmic distance ladder, i.e., the type Ia
supernovae (SNe Ia)~\cite{Riess}.
Therefore, for cosmological studies, an independent direct measurement of $H_{\mathrm{0}}$ with $1\%$ accuracy is of great importance for understanding the aforementioned discrepancy, which may eventually reveal
new physics~\cite{HOLI}.

Recent detections, by advanced laser interferometer gravitational wave observatory (LIGO), of the gravitational wave (GW) signals generated by the mergers of two massive black holes (BHs) opened a new window on the Universe ~\cite{GW150914,GW151226,GW170104}.

In the traditional standard siren approach, the waveform signal from an inspiralling double compact binary can be used to directly measure the
luminosity distance to the source~\cite{Schutz1986}. The calibration uncertainty in strain amplitude is $\lesssim10\%$ for advanced LIGO~\cite{calibration}. Hence, detections of GW together with EM counterpart signals providing the source redshifts, could become excellent cosmological probes~\cite{Zhao2011,Taylor2012}. Binary neutron stars (NS-NS) or neutron star - black hole binaries (NS-BH) are especially promising. They are expected to be seen as kilonovae/mergernovae, short gamma-ray bursts (SGRBs) or fast radio bursts (FRBs)~\cite{counterparts}.

However, the identification of an EM counterpart and associated host galaxy for a GW signal remains challenging given the$\sim 10$ deg$^2$ positional accuracy for GW signals.
Supplementary knowledge might be helpful, like using galaxy catalogs to seek for host galaxy candidates
~\cite{Fan2014,Pozzo2012}.
Knowing the NS equation of state, a tidal correction to the gravitational wave phase in the late-inspiral signal of NS-NS systems~\cite{Messenger2012} or spectral features of the post merger phase~\cite{Messenger2014} can be used to break the mass-redshift degeneracy allowing an estimation of the source redshift and luminosity distance from the GW signal alone.
Another approach is to infer redshifts statistically,  by comparing measured (redshifted) mass distribution of NSs with a universal rest-frame NS mass distribution~\cite{Biesiada2001,Taylor2012}.

Next generation of GW interferometric detectors, like the Einstein Telescope (ET) will broaden the accessible volume of the Universe by three orders of magnitude promising tens to hundreds of thousands of detections per year~\cite{Abernathy2011}, leading to the expectation that many of the sources could be gravitationally lensed. This was discussed by~\cite{JCAP_ET1,JCAP_ET2,JCAP_ET3} with a conclusion that ET should register about 50 -- 100 strongly lensed inspiral events per year, thus providing a considerable catalog of such events during a few years of its successful operation.

The theory of strong gravitational lensing gives the following relationship~\cite{Treu2010}:
\begin{equation}
\Delta t_{i,j} = \frac{D_{\mathrm{\Delta t}}(1+z_{\mathrm{d}})}{c}\Delta \phi_{i,j}, \label{relation}
\end{equation}
where $c$ is the light speed and theoretically GW speed as well. $\Delta t_{i,j}$ is time delay between point images (or two events for GW) $i$ and $j$, $\Delta\phi_{i,j}=[(\boldsymbol{\theta}_i-\boldsymbol{\beta})^2/2-\psi(\boldsymbol{\theta}_i)-(\boldsymbol{\theta}_j-\boldsymbol{\beta})^2/2+\psi(\boldsymbol{\theta}_j)]$
is the difference between Fermat potentials at different image angular positions $\boldsymbol{\theta}_i,\boldsymbol{\theta}_j$, with $\boldsymbol{\beta}$ denoting the source position, and $\psi$ being  the two-dimensional lensing potential determined by the Poisson equation $\nabla^2\psi=2\kappa$, where $\kappa$ is the surface mass density of the lens in units of the critical density $\Sigma_{\mathrm{crit}}=c^2D_{\mathrm{s}}/(4\pi GD_{\mathrm{d}}D_{\mathrm{ds}})$,   $D_{\mathrm{d}}$, $D_{\mathrm{s}}$ and $D_{\mathrm{ds}}$ are  angular diameter distances to the lens (deflector) located at redshift $z_{\mathrm{d}}$, to the source located at redshift $z_{\mathrm{s}}$ and between them, respectively.

The measured time delay between strongly lensed images $\Delta t_{i,j}$ combined with the redshifts of the lens $z_{\mathrm{d}}$ and the source $z_{\mathrm{s}}$, and the Fermat potential difference $\Delta \phi_{i,j}$ determined by lens mass distribution and image positions allow to determine the time-delay distance $D_{\mathrm{\Delta t}}$. This quantity, which is a combination of three angular diameter distances:
\begin{equation}
D_{\mathrm{\Delta t}}=\frac{D_{\mathrm{d}}(z_{\mathrm{d}}) D_{\mathrm{s}}(z_{\mathrm{s}})}{D_{\mathrm{ds}}(z_{\mathrm{d}},z_{\mathrm{s}})},
\end{equation}
contains cosmological information, through the distance-redshift relation.
However, all mass along the light-of-sight (LOS) also contributes to the lens potential with an extra systematic uncertainty at $1\%$ level~\cite{HOLI}.
Therefore, in realistic strong lensing time delay cosmology, we should consider the uncertainties arising from three sources: time delay itself, Fermat potential difference and LOS environment effects.

We show that in the era of third generation ground-based detectors, for lensed GW systems with EM counterparts, the time delay measurements from GW can be quite accurate with ignorable observation error, and the
measurements of Fermat potential differences from EM counterparts can be remarkably improved compared with current lensed quasar systems.
These lensed GW+EM events could thus provide stringent constraints on cosmological parameters, especially the $H_{\mathrm{0}}$ to a very high level.

\begin{center}
{\textbf{ \Large \uppercase{Results}} }
\end{center}

\noindent
{\textbf{Advantages of lensed GW+EM system}}
For the lensed GW and EM systems, we show that both time-delay and Fermat potential difference measurements will be considerably improved compared to the traditional approach to lensed quasars in EM domain~\cite{HOLI}.
Firstly, the time delays measured through GW signals are supposed to be very accurate due to transient nature of double compact object (DCO) merger events ($\sim0.1$s) observed by ground-based GW detectors.
Time delays measured in lensed quasars can achieve at best $3\%$ uncertainty~\cite{Liao2015}.
Secondly, lensed GW signals from such systems are supposed to be associated with
the EM counterparts which are also transient or short events. The kilonovae last only for months, hence the bright transient dominates the host for a relatively short time. This would facilitate identification of the host galaxy of the source in this case.
Acquiring a high-resolution good quality image of the lensed host galaxy before or after the transient event will enable very precise and accurate modeling of the lens.

To understand, quantitatively, the improved accuracy of the lens model with a pure host image (i.e. without the dazzling active galactic nucleus (AGN) images typical in the lensed quasar case), we first used a set of parameters to simulate a typical lensing system, then we added uncertainties to the lensed host image based on the modern quality of Hubble Space Telescope (HST) observation, finally, we tried to recover these parameters
using state-of-the-art lens modelling techniques~\cite{S2013}. This way we estimated the lens modelling precision, i.e., the uncertainty of Fermat potential difference (see the Methods section for details).
We found that the precision or the relative uncertainty of the Fermat potential reconstruction will be improved to $\sim0.6\%$, while the analogous uncertainty in
lensed quasar systems is $\sim3\%$~\cite{HOLI}.

\noindent
{\textbf{Cosmological results}}
To demonstrate the performance of our method, we studied cosmological parameter inference from gravitationally lensed GW and EM signals on a simulated mock data consisting of 10 lensed GW+EM systems. The fiducial cosmology for simulation is flat Lambda Cold Dark Matter model ($\Lambda$CDM) with dimensionless matter density $\varOmega_{\mathrm{M}}=0.3$ and $H_{\mathrm{0}}=70$km$\cdot$s$^{-1}$Mpc$^{-1}$. The data is representative of future observations of lensed GW and EM signals, consisting of lens and source redshifts, accurate time delay measurements, Fermat potential differences with uncertainties and LOS environment uncertainty for each system. The corresponding time delay distances can then be obtained from these data (see the Methods section for details).

\begin{figure}
 \includegraphics[width=12cm,angle=0]{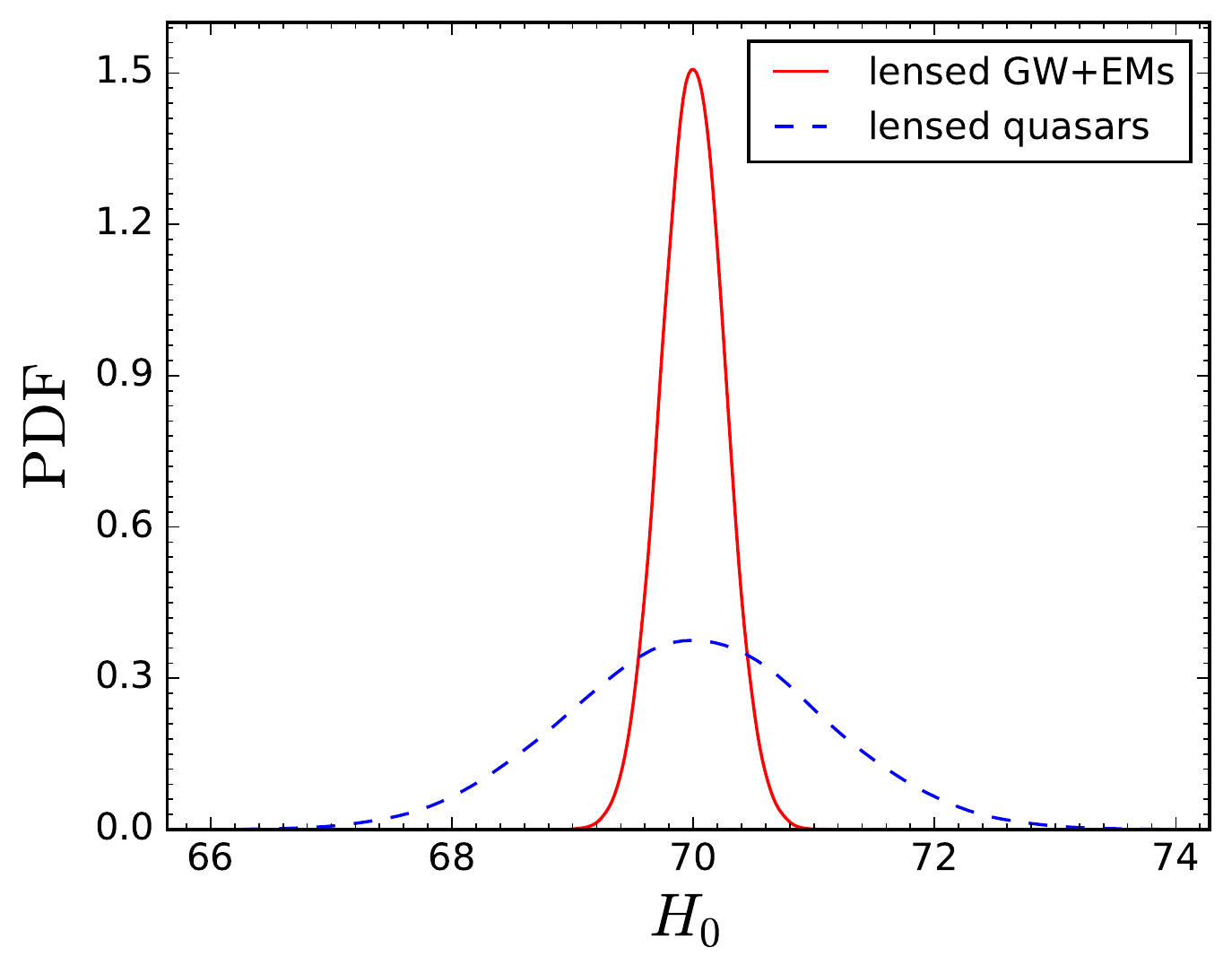}
  \caption{\textbf{Predicted probability distribution function (PDF) of the Hubble constant.} It has been determined from 10 lensed gravitational wave (GW) and electromagnetic (EM) signals assuming flat
   Lambda Cold Dark Matter model ($\Lambda$CDM) and fixed matter density. As a comparison, the case with 10 lensed quasars is also shown. For lensed GW+EM systems,
  the uncertainty of time delay measurement is ignored, the uncertainty of Fermat potential difference is taken as $0.6\%$, the uncertainty of Line of Sight (LOS) environment is $1\%$.
  For lensed quasars, uncertainties of time delay and Fermat potential difference are both taken as $3\%$.
  }\label{H0}
\end{figure}

\begin{figure}
 \includegraphics[width=15cm,angle=0]{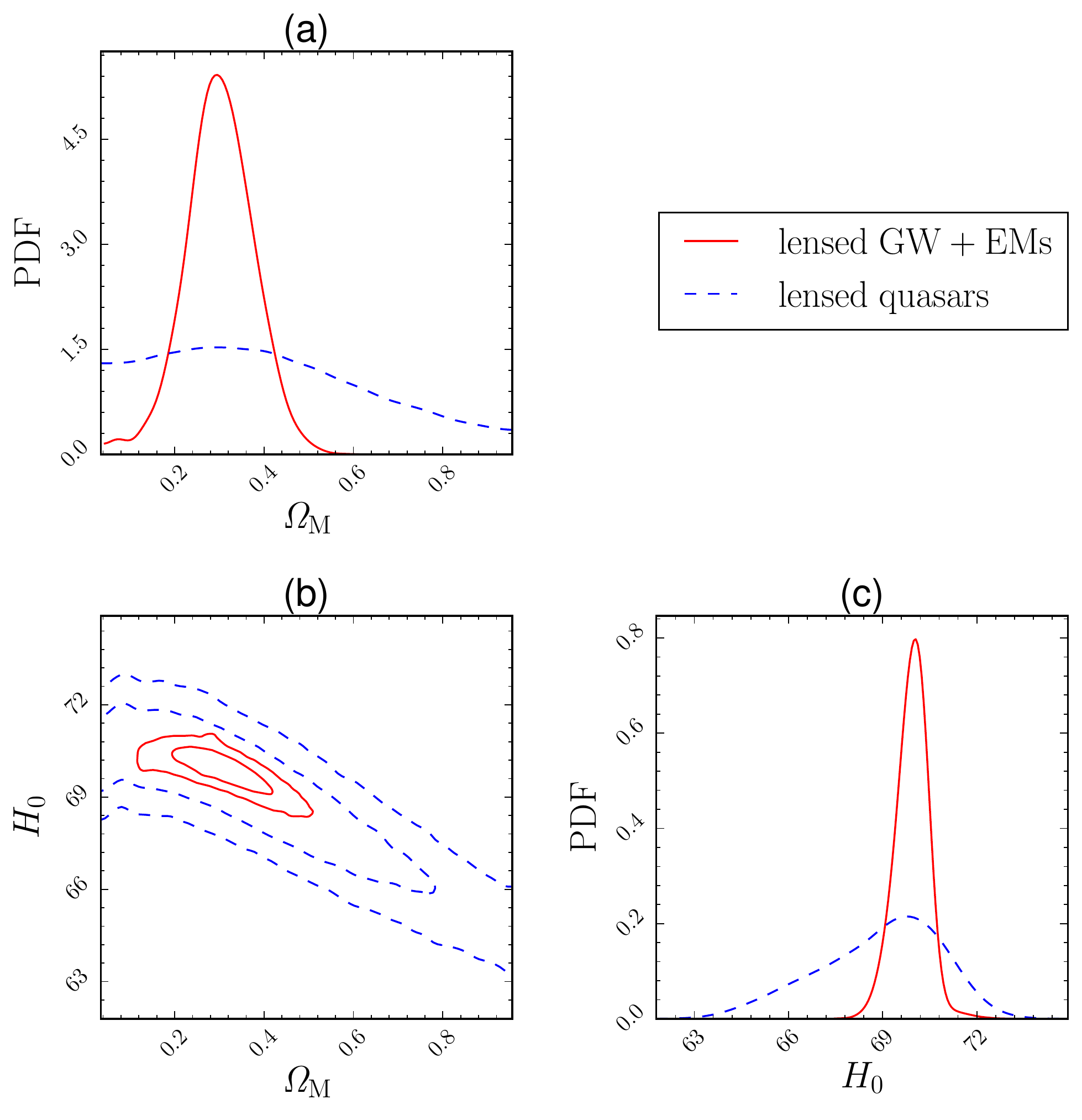}
  \caption{\textbf{Predicted constraints on the parameters in a flat Lambda Cold Dark Matter model ($\Lambda$CDM).} The assumptions are the same as in Fig. \ref{H0}. (a) Marginalized distribution of matter density parameter $\varOmega_{\mathrm{M}}$; (b) 2-D 68\% and 95\% confidence contours for Hubble constant $H_{\mathrm{0}}$ and matter density parameter $\varOmega_{\mathrm{M}}$;
  (c) Marginalized distribution of the Hubble constant $H_{\mathrm{0}}$.
  } \label{lcdm}
\end{figure}

Time delay distance is primarily sensitive to (the inverse of) $H_{\mathrm{0}}$, since $c/H_{\mathrm{0}}$ sets the length scale of the Universe. The dependence on other parameters, such as density parameters or dark energy cosmic equation of state is weaker but can show up when the samples are large or the measurement precision is improved.
Therefore, we first chose a flat $\Lambda$CDM model with matter density $\varOmega_{\mathrm{M}}=0.3$ fixed and we constrained $H_{\mathrm{0}}$ using simulated data. For comparison, we also considered the current state-of-the-art case of lensed quasars~\cite{HOLI}.
Tab.\ref{error} summarizes the uncertainties of three factors contributing to the final uncertainty of time delay distance.
The resulting constraints on $H_{\mathrm{0}}$ in unit of km$\cdot$s$^{-1}$Mpc$^{-1}$ are shown in Fig.\ref{H0}. Lensed GW and EM signals give much more stringent
constraint, the relative uncertainty of $H_{\mathrm{0}}$ being $\sim0.37\%$ in contrast to the lensed quasars observed exclusively in the EM window, having $\sim1.5\%$ relative uncertainty, 4 times larger.
This can be understood because of substantial improvements in time delay and Fermat potential measurements in the multi-messenger systems.
We also considered a flat $\Lambda$CDM universe with the matter density being another free parameter. Fig.\ref{lcdm} shows the confidence contours and marginalized Probability
Distribution Functions (PDFs) of matter density $\varOmega_{\mathrm{M}}$ and $H_{\mathrm{0}}$. The constraining power of lensed GW and EM signals is
also superior to systems observed exclusively in the EM domain. Considering that statistically the precision is inversely proportional to the $\sqrt{N}$, where $N$ is the number of systems, one needs a sample of $\sim160$ time delay systems in a traditional approach in order to get reasonable constraints on parameters other than $H_{\mathrm{0}}$ as in the case. However, future observations of lensed GW and EM signals will enable us to get useful information from just a few such systems. For completeness, we also considered flat $\omega$CDM model where
the coefficient $\omega$ in dark energy equation of state $p = \omega \rho$ is an arbitrary constant and an
open $\Lambda$CDM model where the spatial curvature $\varOmega_{\mathrm{k}}$ of the Universe is not fixed as vanishing. The results are shown in Tab.\ref{table}.

\begin{table}
\begin{center}
\begin{tabular}{lccc}
\hline\hline
 & $\delta\Delta t$ & $\delta\Delta\psi$ &  $\delta LOS$\\
\hline
lensed GW+EM &0\% &0.6\% &1\% \\
lensed quasar &3\% & 3\% &1\%  \\
\hline\hline
\end{tabular}
\end{center}
\caption{\textbf{Relative uncertainties of three factors contributing to the accuracy of time delay distance measurement.}
$\delta\Delta t$, $\delta\Delta\psi$, $\delta LOS$ correspond to time delay, Fermat potential difference and light-of-sight environment, respectively.
We show the case for lensed gravitational wave (GW)+ electromagnetic (EM) signals compared with
standard technique in the EM domain using lensed quasars.
}\label{error}
\end{table}

\begin{table*}
 \begin{tabular}{lcccccccccccc}
  \hline
  &Flat $\Lambda$CDM ($\varOmega_{\mathrm{M}}$ fixed) & & \multicolumn{2}{c}{Flat $\Lambda$CDM} & & \multicolumn{3}{c}{Flat $\omega$CDM}
& &\multicolumn{3}{c}{Open $\Lambda$CDM} \\
  & $H_{\mathrm{0}}$ & & $H_{\mathrm{0}}$ & $\varOmega_{\mathrm{M}}$  & & $H_{\mathrm{0}}$ & $\varOmega_{\mathrm{M}}$  & $w$
& & $H_{\mathrm{0}}$ & $\varOmega_{\mathrm{M}}$ & $\varOmega_{\mathrm{k}}$  \\
  \hline
 Uncertainty & $0.37\%$& & $0.68\%$ &
$27\%$  && $2.2\%$ & $36\%$
& $25\%$ & & $1\%$ &
$38\%$ & $\pm 0.18$ \\
  \hline
 \end{tabular}
 \caption{\textbf{The average constraining power of 10 lensed
 gravitational wave (GW)+electromagnetic (EM) systems.} We concerns cosmological parameters in different scenarios: flat Lambda Cold Dark Matter (Flat $\Lambda$CDM)
 with or without dimensionless matter density $\varOmega_{\mathrm{M}}$ fixed, flat $\omega$CDM where the dark energy equation of state $\omega$ is a free parameter, and
 open $\Lambda$CDM where cosmic curvature $\varOmega_{\mathrm{k}}$ is a free parameter.
 For the same number of lensed quasars, the power is weaker by a factor of $\sim4$ according to the uncertainty propagation using Eq.\ref{relation} and Tab.\ref{error}.
}\label{table}
\end{table*}

\newpage

\begin{center}
{\textbf{ \Large \uppercase{Discussion}} }
\end{center}
Let us compare cosmological applications of strong lensing discussed in the literature.
In the EM window, strong lensing time delays of AGNs in quasars plus the host galaxy observation are known as a cosmological tool~\cite{Refsdal1964} (see also~\cite{Treu2016}). Recently this technique enabled the determination of the Hubble constant with a few percent precision~\cite{HOLI}.
The upcoming Large Synoptic Survey Telescope (LSST) will enable the first long baseline multi-epoch observational campaign on several thousand lensed quasars~\cite{OM10}.
The strong lens time delay challenge program (TDC)~\cite{Liao2015} has proven
that the LSST will yield $\sim400$ quasar-elliptical galaxy
systems with well-measured time-delay light curves, with $\Delta t_{i,j}$ measurements up to precision $\sim3\%$ including systematics. On the other hand,
current 
high resolution imaging of the host combined with spectroscopic observations of stellar kinematics of the lens galaxy could give similar $\sim3\%$ uncertainty (including the systematics) concerning the Fermat potential~\cite{HOLI}.

Lensing of pure GW signals has already been discussed in the literature~\cite{Wang1996,Nakamura1998, Takahashi2003,Cao2014}. In the context of
Laser Interferometer Space Antenna (LISA) interferometric detector in space, weak lensing causes significant uncertainties of luminosity distance measurements (see discussion and references in \cite{Hogan2009}). Strong lensing of LISA target sources (supermassive BHs) has been discussed in \cite{Sereno2010}, and \cite{Sereno2011} proposed to use the statistics of strongly lensed sources or the time-delay measurements of lensed GW signals to constrain cosmological parameters without identifying the EM counterparts. It was shown that these approaches could constrain the Hubble constant with $\sim10\%$ precision. Note that the inspiral signal from supermassive BHs involves much longer time scales of event time and waveform variations than in the transient sources recorded by ground-based detectors that have event times $\sim0.1$s, implying the ground-based detectors would get quite accurate time delay measurements in a waveform independent way.

In comparison to standard techniques, our method has the following advantages. First of all, lensed GW signal detection coordinated with EM searches 
(possibly at different wavelengths) 
would facilitate source identification. Even if EM transients would be missed, gravitationally lensed systems could be searched through catalogs from large synoptic surveys within the broad location band provided by GW detector.
The proposed method of cosmographic inference is waveform independent in its principle. It is not necessary to disentangle fine details of the waveform leading to precise measurements of chirp masses or luminosity distances. One only needs to uncover the lensed nature of two GW signals by establishing that they differ only by amplitude having the same duration, frequency drift and rate of change of the amplitude.
Even though we emphasize that precise waveform analysis is not crucial to our method, yet possible estimates of source luminosity distance would provide another boundary condition facilitating identification of strongly lensed system in the EM domain. Time delay determination from lensed GW signal would reach an unprecedented accuracy $\sim 0.1$s from the detection pipeline or even by many orders of magnitude higher if the details of the waveform are analyzed, e.g. the moment of final coalescence can be determined with $\sim 10^{-4}$ms accuracy.
Such accurate measurements of lensing time delays can become a milestone in precision cosmology.

Gravitationally lensed systems seen in GW and EM signals could be used to test modified theories of gravity~\cite{Fan2017,Collett2017}. They can also serve as consistency tests for gravitational lensing studies in EM domain. 
Besides, accurate time delay measurements can be applied to studying galaxy structure, for example, the mass density slope of elliptical galaxies and its evolution with redshift, and dark matter substructure in galaxy-scale halos~\cite{darkmatter}.
Although the method we propose may be limited by the number of detections of lensed GW+EM systems,
we look forward to seeing these systems detected and applied to cosmological and astrophysical studies in the near future.

\begin{center}
{\textbf{ \Large \uppercase{Methods}} }
\end{center}

\noindent
{\textbf{Mock data generation}}
We generated the mock data taking into account the uncertainty levels reported in Tab.\ref{error}.
The data consisted of simulated values comprising the following quantities: redshift of the lens and of the source (assumed to be accurate), strong lensing time delays (assumed to be measured accurately), Fermat potential difference together with its uncertainty inferred from images of lensed host galaxy, and an extra uncertainty of the inferred time delay distance caused by perturbers along the line of sight.

The choice of redshifts of the source and deflector may affect the result of cosmological constraints, thus they must be selected carefully in order to represent fairly the constraining power of randomly chosen 10 strong lensing systems.
Therefore, we generated a set of redshifts of sources and deflectors, based on the redshift probability distribution functions (PDFs)
calculated by ~\cite{JCAP_ET2,JCAP_ET3}. These PDFs were obtained in the following way:
Firstly, taking into account full population of DCOs, i.e., NS-NS, NS-BH, and BH-BH binaries with their intrinsic merger rates at different redshifts calculated with the population synthesis code {\tt StarTrack}~\cite{StarTrack}, and the expected sensitivity of
ET, the number of yearly detected GW events was predicted (Table 1 of ~\cite{JCAP_ET2}).
Secondly, the probability of each GW signal from inspiralling DCO lensed by early-type galaxies with lensed signals magnified sufficiently to be detected by ET was calculated.
The deflectors were assumed as Singular Isothermal Spheres (SIS) with the velocity dispersions following Schechter distribution.
Lastly, summing all the DCO merging systems together,
the total number of lensed events registered by the ET per year was predicted. This prediction is accompanied by
the redshift PDF (see Fig.~2 in Ding et al.~\cite{JCAP_ET3}), which
enables us to randomly generate the samples of redshifts of the sources and deflectors.
We used the standard scenario of NS-NS and NS-BH systems merging history with "low-end" metallicity evolution \cite{StarTrack} to randomly generate 300 systems with lens and
source redshifts.

Then, we assigned time delays to each system, typically several tens of days. Time delays depend on the redshifts $z_{\mathrm{d}}$ and $z_{\mathrm{s}}$, velocity dispersion of the lens and the random relative source position on the source plane. We used the parameters according to OM10 catalog made by Oguri and Marshall~\cite{OM10}.
Using Eq.\ref{relation} and knowing redshifts $z_{\mathrm{d}}$ and $z_{\mathrm{s}}$, we calculated theoretical time delay distance $D_{\mathrm{\Delta t}}$ based on fiducial cosmological model  i.e. flat $\Lambda$CDM, and also flat $\omega$CDM or open $\Lambda$CDM, respectively.
Next, we calculated theoretical Fermat potential difference between two image positions and we added $0.6\%$ uncertainties
to it. The values obtained this way were treated as the simulated Fermat potential difference data.

In the last step, since in addition to the lens galaxy mass distribution, the structures along the
line of sight also affect the time delay distance~\cite{Collett2013}, i.e., the external masses and voids
make additional focussing and defocussing of the light rays, we considered the extra uncertainty from the LOS contamination.
If the effects of LOS perturbers are small, they can be approximated by an external convergence term in the lens plane, $\kappa_{\mathrm{ext}}$.
The true $D_{\mathrm{\Delta t}}$ is then related to the modeled one by $D_{\mathrm{\Delta t}}=D_{\mathrm{\Delta t}}^{\mathrm{model}}/(1-\kappa_{\mathrm{ext}})$.
One can estimate $\kappa_{\mathrm{ext}}$ from galaxy counts~\cite{galaxycount} and tracing rays through the Millennium Simulation~\cite{MS}.
We assumed the corresponding uncertainty as $1\%$ of the inferred time delay distance $D_{\mathrm{\Delta t}}$ from Eq.\ref{relation} as suggested by the
$H_{\mathrm{0}}$ Lenses in COSMORAIL's Wellspring program (H0LiCOW)~\cite{HOLI}, where COSMOGRAIL stands for the COSmological MOnitoring of GRAvItational Lenses program~\cite{COSMOGRAIL}.

\noindent
{\textbf{Lensed GW and EM signals}}
Elaboration of GW detector data analysis pipeline for identifying lensed GW signals is an ongoing study undertaken by a few groups.
It has not been a top issue for advanced LIGO since the probability of observing such events in this generation of detectors is
very small~\cite{Wang1996,Dai2017}. Now, however it is becoming important partly because of looking toward to a new generation of detectors in which such events
could be registered more frequently and partly because of the benefits stemming from such detections (e.g. ~\cite{Fan2017, Baker2017} or discussions in ~\cite{JCAP_ET2, JCAP_ET3}).

 The signature of lensed GW signals would be that they differ only by amplitude having the same duration, frequency drift, rate of change of the amplitude (i.e. the chirp) and come from the same location strip on the sky. The amplitude scale of the signal could also be affected by the detector's orientation factor changing between the arrivals of lensed signals due to rotation of the Earth, but this could be accounted for once the time delay is known.
Moreover, this would affect only the determination of flux ratios between images, which are not an important part of our method.
In any case, true benefits would come from the multi-messenger nature of such an event~\cite{Baker2017}. Therefore the crucial part is a cross-confirming procedure in both GW and EM domains.

We cannot be more quantitative here because appropriate pipelines for coordinated searches of lensed events in EM and GW domains have not yet been constructed or validated.
Attractiveness of such detections, supported among others by the findings we report in this letter, will certainly boost the development of such pipelines. However, we outline below, the main steps of a realistic approach.
A single detection in one domain should trigger a coordinated search in the data from the other one, e.g. if GW data analysis provides a pair of events suspected of being lenses, this should trigger a search for lensed (repeated) EM transients in the sky location strip of GW source.
Conversely, if a lensed kilonova event is observed in a large survey telescope, this should trigger confirmation searches in the GW signal database for coherent waveforms and time delay between them consistent with EM signal. Let us note that a rough estimate of time delay would be possible from kilonovae light curves in multiple images. The demand that both GW and EM signals are lensed and arrive with the same time delay is a considerable restriction imposed on possible EM counterparts of GWs.
After confirmation that two GW signals come from the same source and the counterpart is a kilonova~\cite{Kilonovae}, one can take the value of
time between these two GW transients as representing the accurate lensing time delay with uncertainty smaller than or comparable to the event time $\sim0.1$s (see~\cite{timescales} for estimations of different event time scales).

\noindent
{\textbf{Fermat potential improvements}}
For traditional quasar system, both lens model and the Fermat potentials are recovered from lensed host galaxy image by
extracting the AGN component. This is done using a nearby star's Point Spread Function (PSF) or by adopting an iterative modeling process which can accurately recover the PSF for real observations~\cite{Chen2016,Wong2017,S2013,HOLIV}.
Unfortunately, these operations cannot totally eliminate systematic errors, especially in the central part of AGN, because of difficulties associated with the following three aspects. First, due to huge intensity of AGN, even a tiny mismatch when extracting the AGN images as the scaled PSFs, would lead to a non-negligible discrepancy.
Second, to avoid the saturation of the Charge-coupled Device (CCD) of space telescope like HST,
the central AGN area is taken with short exposure time while the other region is taken with long exposure time.
Therefore, the pixels in the central AGN area have large uncertainties,
and quite rough, which introduce a severe bias.
Lastly, the dithering and drizzling operations would slightly (but non-negligibly) shift the light distribution in the central
AGN which make the lens modeling in this area even harder.
In order to test the fidelity of lens modeling techniques, Ding et al.~\cite{Ding2017} carried out a simulation exercise.
They found, even if the perfect PSF is given, a significant residual in the central AGN area is still inevitable~\cite{Ding2017} (Figure 4-b, therein).
Fortunately, one does not encounter these difficulties while studying the lensed GW+EM events since these systems do not possess the bright point images.
In principle, lens modeling and inference of the Fermat potentials from lensed GW+EM system would be much more precise and accurate.

To compare the precision of lens modeling between AGN and GW+EM systems directly,
we simulated two sets of realistic lensed images with and without the AGN, based on the current lensing project H0LiCOW~\cite{HOLI}. We refer to section 3 of Ding et al.~\cite{Ding2017} for a detailed description of such a simulation approach.
During the simulation, exposure time and noise level were set to values based on deep HST observations.
In order to assess the accuracy of the Fermat potential recovery, in our simulations we treated the parameters
in an elliptically symmetric power-law lens model, for example the radial slope, as free parameters to be inferred from observations. We found that the effect of bright PSFs influences the uncertainties of
these parameters by at least a factor of five. Given that the current lens modeling technique recovers the Fermat potential at $3\%$ uncertainty level~\cite{HOLI}, we conclude that with gravitationally lensed GW+EM signals, the lens modeling would yield the Fermat potential with $0.6\%$ uncertainty, though this number depends on the real observing conditions.

Let us note that, for a lensed quasar observation with relatively large uncertainties, we may need to choose a specific lens mass model during the lens modelling,
for example, the power-law or a composite model with a baryonic component and a Navarro-Frenk-White (NFW) dark matter halo.
When the observation is precise in the lensed
GW+EM case, i.e., the pure host without bright PSF contamination, we can make a better decision of the fiducial lens model, and this will decrease the systematical bias.

\noindent
{\textbf{Statistical analysis.}}
A particular single strong lensing system possesses its own sensitivity to cosmological parameters due to its specific combination of lens and source redshifts. In order to show the representative average constraining power from 10 such systems, we randomly selected 30 datasets each containing 10 strong lensing systems from the 300 systems mentioned above.
Then we propagated the relative uncertainties of the Fermat potential difference and the line of sight contamination to the relative uncertainty of $D_{\mathrm{\Delta t}}$, and then to the relative uncertainties of cosmological parameters on which it depends: $(\delta\Delta\psi,\delta \Delta t, \delta LOS)\sim \delta D_{\mathrm{\Delta t}}\sim (\delta H_{\mathrm{0}}, \delta\Omega_{\mathrm{M}},\omega,\varOmega_{\mathrm{k}})$.
The relative time delay uncertainty was assumed $\delta \Delta t=0$ for lensed GW and EM signals, while for quasars -- studied for comparison -- it was assumed at the level of $3\%$. We performed Markov Chain Monte Carlo (MCMC) minimizations using Python module PyMC applied to the $\chi^2$ objective function:
\begin{equation}
\chi^2=\sum\limits_{i=1}^{10}{(D_{\mathrm{\Delta t},i}^{\mathrm{th}}(z_{\mathrm{d},i},z_{\mathrm{s},i},H_{\mathrm{0}},\varOmega_{\mathrm{M}},\omega,\varOmega_{\mathrm{k}})-D_{\mathrm{\Delta t},i}^{\mathrm{sim}})^2/\sigma^2_{ D_{\mathrm{\Delta t},i}}},
\end{equation}
where $D_{\mathrm{\Delta t}}^{\mathrm{th}}$ is the time delay distance calculated in the assumed cosmological model, while $D_{\mathrm{\Delta t}}^{\mathrm{sim}}$ is the corresponding distance inferred from simulated Fermat potential difference with extra LOS uncertainty considered and its uncertainty is $\sigma_{D_{\mathrm{\Delta t},i}} = \delta D_{\mathrm{\Delta t},i} D_{\mathrm{\Delta t},i}$.  Parameters were sampled from ranges $H_{\mathrm{0}} \in [0, 150], \varOmega_{\mathrm{M}} \in [0, 1.5], \omega \in [-2,0], \varOmega_{\mathrm{k}} \in [-1, 1]$.

For each dataset we obtained the marginalized distributions for each cosmological parameter.
From the resulting distributions we calculated respective $1\sigma$ uncertainties and after averaging them over 30 datasets we reported the results in Table.\ref{table}. We plotted the PDFs and confidence contours of cosmological parameters recovered from one of the datasets in Fig.\ref{H0} and Fig.\ref{lcdm}.

\noindent
{\textbf{Data availability.}} The data that support the findings of this study are available from the corresponding author upon request.

\begin{addendum}
\item [Acknowledgements] We thank C. Messenger and T. Collett for helpful discussions. This work was supported by Natural Science Foundation of China under Grants No. 11633001,  No. 11673008, No. 11603015. X. Fan was also supported by Newton International Fellowship Alumni Follow on Funding. M. Biesiada gratefully acknowledges hospitality and support of Beijing Normal University where a part of this work was performed.

\item[Author contributions]
K. Liao contributed in proposing the idea using lensed GW and EM systems to infer Hubble constant precisely. He also contributed in
calculations and original paper writing.
X.-L. Fan contributed in proposing the measurement strategy and the idea using lensed GW and EM systems to study cosmology, as well as in organizing the research.
X. Ding contributed in Fermat potential simulations.
M. Biesiada contributed in discussing and proposing the new ideas, comparing the work with relevant literature and improving the quality of the paper.
Z.-H. Zhu contributed in discussing the new ideas and organizing the research.

\item[Competing interests]The authors declare no competing financial interests.

\end{addendum}


\end{document}